\documentclass[pra,twocolumn,amsmath,amssymb,groupedaddress,superscriptaddress]{revtex4}		
\usepackage{graphics,xcolor}
\usepackage{amsmath}
\usepackage{float,graphicx}
\usepackage{subfigure}
\usepackage{epstopdf}
\usepackage{soul}
\usepackage{dsfont}
\usepackage{booktabs}
\usepackage[normalem]{ulem}
\usepackage{footnote}
\usepackage{mathtools}
\usepackage{tikz}	

\usepackage[linkcolor=black, urlcolor=black, colorlinks=true, citecolor=cyan]{hyperref}

\newcommand{\ie}{i.\,e.~}

\newcommand{\eg}{e.\,g.~}

\newcommand{\cf}{c.\,f.~}

\newcommand{\wrt}{w.\,r.\,t.~}

\def\ga{\alpha}
\def\gb{\beta}
\def\ggg{\gamma}

\def\gd{\delta}
\def\gD{\Delta}
\def\ge{\varepsilon}

\def\gt{\vartheta}
\def\gT{\theta}
\def\gk{\kappa}

\def\gf{\varphi}

\def\cG{\mathcal{G}}

\def\cN{\mathcal{N}}

\newcommand{\unit}{\mathds{1}}								
\newcommand{\bra}[1]{\langle #1 \vert} 					
\newcommand{\ket}[1]{\vert #1 \rangle} 					
\newcommand{\ip}[2]{\langle #1 \vert #2 \rangle}    		
\newcommand{\bmat}[1]{\left(\begin{array}{#1}}				
\newcommand{\emat}{\end{array}\right)}						
\newcommand{\cas}[4]{\left\{ \begin{array}{lcl} #1 & for & #2 \\ #3 & for & #4 \end{array} \right.}

\newcommand{\real}[1]{\mathfrak{Re}\left\{ #1 \right\}}	
\newcommand{\order}[1]{\mathcal{O}\left( #1 \right)}		

\newcommand{\p}[3]{\prod_{#1=#2}^{#3}}						
\newcommand{\s}[3]{\sum_{#1=#2}^{#3}}						

\def\dag{^{\dagger}}										

\def\w/{\textmd{with}}										
\def\NN{\nonumber}											

\PassOptionsToPackage{smaller}{acronym}
  \usepackage[nolist]{acronym} 

\begin{document}
\title{Avoiding local minima in variational quantum eigensolvers\\with the natural gradient optimizer}
\author{David Wierichs\footnote{wierichs@thp.uni-koeln.de}}
\affiliation{Institute for Theoretical Physics, University of Cologne, Germany}
\author{Christian Gogolin}
\affiliation{Institute for Theoretical Physics, University of Cologne, Germany}
\affiliation{Covestro Deutschland AG, Kaiser Wilhelm Allee 60, 51373 Leverkusen, Germany}
\author{Michael Kastoryano}
\affiliation{Institute for Theoretical Physics, University of Cologne, Germany}
\affiliation{Amazon Quantum Solutions Lab, Seattle, Washington 98170, USA}
\affiliation{AWS Center for Quantum Computing, Pasadena, California 91125, USA}

\begin{abstract} 
We compare the \acs{BFGS} optimizer, \acs{ADAM} and \ac{NatGrad} in the context of \acp{VQE}.
We systematically analyze their performance on the \acs{QAOA} ansatz for the \ac{TFIM} as well as on overparametrized circuits with the ability to break the symmetry of the Hamiltonian. 
The \acs{BFGS} algorithm is frequently unable to find a global minimum for systems beyond about 20 spins and \acs{ADAM} easily gets trapped in local minima. On the other hand, \acs{NatGrad} shows stable performance on all considered system sizes, albeit at a significantly higher cost per epoch.
In sharp contrast to most classical gradient based learning, the performance of all optimizers is found to decrease upon seemingly benign overparametrization of the ansatz class, with \acs{BFGS} and \acs{ADAM} failing more often and more severely than \ac{NatGrad}. 

Additional tests for the Heisenberg \acs{XXZ} model corroborate the accuracy problems of \acs{BFGS} in high dimensions, but they reveal some shortcomings of \acs{NatGrad} as well. 
Our results suggest that great care needs to be taken in the choice of gradient based optimizers and the parametrization for \acp{VQE}.

\end{abstract}
\maketitle
\begin{acronym}[UMLX]
	\acro{VQE}{Variational Quantum Eigensolver}
	\acro{QAOA}{Quantum Approximate Optimization Algorithm}
	\acro{QAOa}{Quantum Alternating Operator ansatz}
	\acro{TFIM}{Transverse Field Ising Model}
	\acro{XXZ}{}
    \acro{XXZM}{Heisenberg \acs{XXZ} Model}
	\acro{BFGS}{Broyden-Fletcher-Goldfarb-Shanno}
	\acro{ADAM}{Adaptive Moment Estimation}
	\acro{RBM}{Restricted Boltzmann Machine}
	\acro{PBC}{periodic boundary conditions}
	\acro{SPSA}{Simultaneous Perturbation Stochastic Approximation}
    \acro{NISQ}{Noisy Intermediate-Scale Quantum}
    \acro{NatGrad}{Natural Gradient Descent}
    \acro{EVQE}{}
    \acro{FLOP}{}
    \acro{SWAP}{}   
    \acro{CNOT}{}
    \acro{ADAPT-VQE}{}
    \acro{DIRECT}{}
    \acro{RMSprop}{}
    \acro{AdaGrad}{}
\end{acronym}

\acresetall 

\section{Introduction}
Variational quantum algorithms such as the \ac{VQE} or the \ac{QAOA} \cite{Farhi_Goldstone_14} have received a lot of attention of late. They are promising candidates for gaining a quantum advantage already with \ac{NISQ} computers in areas such as quantum chemistry \cite{Cao_Aspuru-Guzik_19}, condensed matter simulations \cite{Smith_Knolle_19}, and discrete optimization tasks \cite{Zhou_Lukin_18}.
A major open problem is that of finding good classical optimizers which are able to guide such hybrid quantum-classical algorithms to desirable minima and to do that with the smallest possible number of calls to a quantum computer backend. 
In classical machine learning, the \ac{ADAM} \cite{Kingma_Ba_14} optimizer is among the most widely used and recommended algorithms \cite{Karpathy_17, Ruder_16}, and has been one of the most important enablers of progress in deep learning  in recent years. Such an accurate and versatile optimizer for quantum variational algorithms is yet to be found.

We are here mostly interested in variational algorithms for quantum many-body problems.
To make progress towards finding an efficient and reliably optimizer for this domain, we concentrate on cost functions derived from typical quantum many-body Hamiltonians such as the \ac{TFIM} and the \ac{XXZM} for two reasons:
First, their system size can be varied allowing us to systematically study scaling effects.
Second, for \textit{integrable} systems such as the \ac{TFIM} the exact ground states are known and it is possible to construct ansatz classes for \ac{VQE} circuits that provably contain the global minimum and can be simulated more efficiently.
Such systems thus allow us to distinguish between the  performance of the optimizers and the expressiveness of the ansatz.

As a first result we show that the commonly used optimization strategies \ac{ADAM} \cite{Ostaszewski_Benedetti_19} and \ac{BFGS} \cite{Broyden_70, Fletcher_70, Goldfarb_70,Shanno_70,Guerreschi_Smelyanskiy_17, Mbeng_Santoro_19, Wang_Rieffel_19, Grimsley_Mayhall_19, Romero_Aspuru-Guzik_17, Gard_Barnes_20} both run into convergence problems when the system size of a \ac{VQE} is increased.
This happens already for system sizes within the reach of current and near future \ac{NISQ} devices, which underlines the importance to a systematic search for suitable optimization strategies.
The \ac{BFGS} algorithm fails systematically for bigger systems above about 20 spins in the \ac{TFIM} corresponding to $20$ variational parameters. 
The performance of \ac{ADAM} is shown to depend strongly on the learning rate via multiple effects and the number of epochs required for convergence increases fast with the problem size.
Convergence can be improved but only with an expensive fine-tuning of the hyperparameters.

We then study the performance of an optimization strategy known as the Quantum Natural Gradient or \acs{NatGrad} \cite{Stokes_Carleo_19, Amari_98, Harrow_Napp_19} and introduce Tikhonov regularization to the classical processing step in the \ac{VQE} \cite{Martens_Sutskever_12}. 
We find that \ac{NatGrad} regularized in this way does consistently find a global optimum for the largest system sizes we test (40 qubits) and requires significantly fewer epochs to do so than \ac{ADAM} (in the cases where \ac{ADAM} converges at all). 

This is in sharp contrast to the usually very good performance of the \ac{ADAM} optimizer and related (stochastic) gradient descend based techniques in the optimization of classical neural networks.
A possible explanation for this good performance in usually overparametrized settings is the following:
For common activation functions and random initialization, increasing overparametrization tends to transform local minima into saddle points \cite{Livni_Shamir_14, Li_Sun_18}.
The optimizer then mainly needs to follow a deep and narrow valley with comparably flat bottom to find a global minimum.
The \ac{ADAM} optimizer is perfectly suitable to pursue this path as it has per-parameter learning rates that also take into account the average of recent updates.
In this way it avoids side-to-side oscillations in the valley and can build up momentum to slide down the relatively flat bottom of the valley. 

The energy landscapes of typical variational quantum algorithms however look very different.
First, having deep and wide circuits with many parametrized gates is prohibitive on \ac{NISQ} computers, which excludes overparametrization as a tool to make the variational space more accessible.
Second, the variational parameters usually feed into exponentially generated gates and thus the cost function is a combination of trigonometric functions of the parameters.
It appears that \ac{NatGrad} is able to effectively use the information about the ansatz class to navigate the resulting energy landscape with many local minima.
Third, it is known that large parts of the parameter space form so-called barren plateaus with very small gradients \cite{McClean_Neven_18}.
A random initialization of the parameters in reasonably deep \ac{VQE}s is thus almost certainly going to leave one stuck in such a plateau.
Of course this also implies that one must prevent the optimizer from jumping to a random location in parameter space during optimization.
This can be achieved in \ac{NatGrad} by inhibiting unsuitably large steps by means of Tikhonov regularization. 
Finally, due to the small number of variational parameters in VQEs, the added (classical) computational cost of inverting the Fubini-Study metric is neglegible as compared to the cost of sampling from the quantum backend. 
This fact, combined with the highly correlated nature of the learning landscape in quantum many-body problems \cite{Park_Kastoryano_19}, might render second-order methods such as \ac{NatGrad} more amenable to quantum than to classical settings, where samples are cheap, but there are many variational parameters.

Our second set of results concerns the effect of overparametrization in \acp{VQE}.
We study the impact of adding redundant layers to the ideal circuit ansatz.
Not only does this overparametrization not improve the performance, it actually appears to make finding the optimum significantly harder.
The \ac{BFGS} algorithm but also the \ac{ADAM} optimizer, designed to thrive on additional degrees of freedom, fail frequently in this setting.
This cannot easily be mitigated by increasing the iteration budget and reducing the learning rate of the \ac{ADAM} optimizer.
While also affected, \ac{NatGrad} shows much higher resilience against this effect, compensating its higher per-epoch cost with a higher chance to succeed.

In order to generalize our results, we consider the \ac{XXZM} together with the Trotterized time evolution operator as circuit ansatz. 
Indeed we find \ac{BFGS} to experience the same difficulties in high-dimensional parameter spaces and \ac{ADAM} to exhibit a similar behaviour of the required number of epochs as for the \ac{TFIM}. 
The performance of \ac{NatGrad} however is not as reliable for this model as it shows very flat intermediate optimization curves which obscur the distinction between challenging phases in the computation and convergence to local minima. 
A detailed investigation of the origin of the deviating behaviour and potential improvements of \ac{NatGrad} will be subject of future work. 

\section{Main results}
In this section, we state and assess the main numerical results of the paper. 
For a detailed description of the optimizers and circuit models, see the Methods section (sec.~\ref{sec:methods}). 

\begin{figure}[h!]
\flushleft
\includegraphics[width=0.48\textwidth]{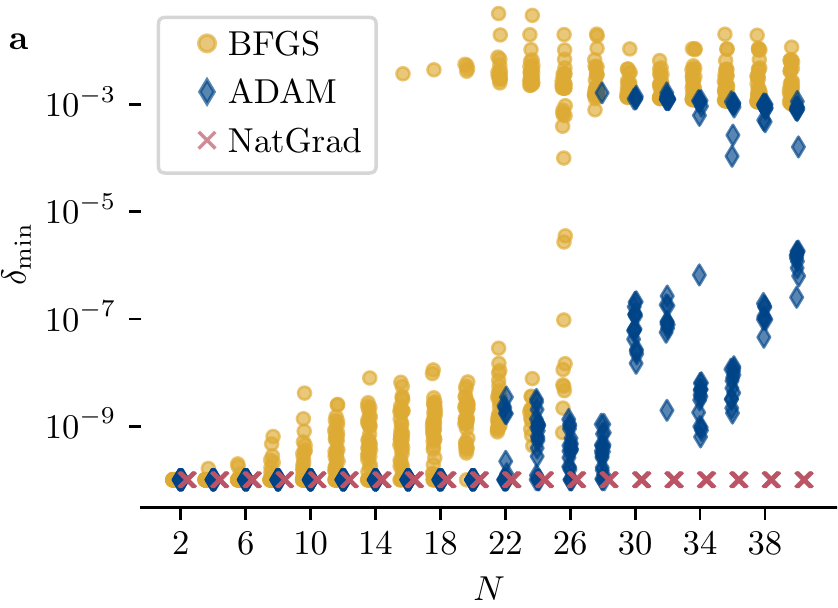}

\vspace{0.25cm}
\includegraphics[width=0.48\textwidth]{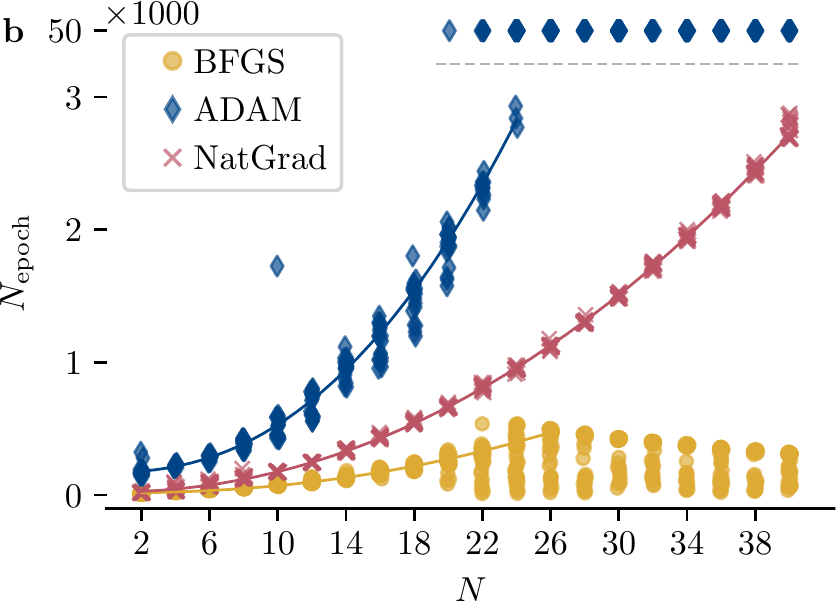}
\caption{Relative error $\gd_\text{min}$ and epoch count $N_\text{epoch}$ for the three optimizers with random initialisation for the \ac{QAOA} circuit with $n=N$ variational parameters. 
The \ac{ADAM} optimizer is chosen with a learning rate of $\eta=0.06$. 
 (a) NatGrad reaches the ground state for all instances and all system sizes, while BFGS and ADAM start systematically getting stuck in local minima beyond a system size of $N=20$. (b) The monomial fits to the mean number of epochs to global minimization yield the scalings $N^{2.1}$ (\ac{BFGS}), $N^{2.3}$ (\ac{ADAM}) and $N^{2.1}$ (\ac{NatGrad}). ADAM experiences a transition around $N=22$ qubits, where the number of epochs to convergence jumps by an order of magnitude. 
\label{fig:tfi_nondist_prec}}
\end{figure}

\subsection{QAOA circuits for the TFIM}\label{sec:res_tfi_qaoa}
We start our numerical investigation with the \ac{QAOA} circuit for the \acs{TFIM} on $N$ qubits with a depth of $p=N/2$ blocks and analyze the \textit{accuracy, speed and stability} of all three optimizers \ac{BFGS}, \ac{ADAM} and \ac{NatGrad} (see sec.~\ref{sec:qaoa} for the ansatz and \ref{sec:tfi} for the model).
These circuits with $n=N$ parameters are sufficiently expressive to contain the ground state and respect the symmetries of the Hamiltonian. 
For each system size we sample $20$ points in parameter space and initialize each optimizer at these positions.
This leads to statistically distributed performances of the algorithms and as we perform exact simulations without sampling and noise it is the only source of stochasticity.
The minimal relative error $\gd_\text{min}$ and the number of required epochs for each initial point and optimizer are shown in fig.~\ref{fig:tfi_nondist_prec}.

Before we analyze the results, recall that the optimization problem can be solved exactly, \ie the ansatz contains the true ground state.
This enables us to identify optimization results with precisions $\gd_\text{min}\geq 10^{-3}$ as local minima and we consider them to be unsuccessful.
In practical applications the precision reached in both local and global minima would be much lower and in particular results with $\gd\approx 10^{-10}$ are unreasonable to measure in quantum machines.
This choice of benchmark is made in order to clearly reveal intrinsic features of the optimizers. 
For realistic applications, a systematic study of noise needs to be taken into account as well.

Our first observation is that the \ac{BFGS} optimizer systematically fails to converge for systems sizes larger than $N=20$.
For small system sizes, however,  it reaches a global minimum in the smallest number of iterations and at low cost per epoch (see tab.~\ref{tab:optimizercost}).
The runs of \ac{BFGS} interrupted at a $\gd<10^{-6}$ level could be improved to reach the goal of $\gd=10^{-10}$ by tuning the interrupt criterion. Therefore, these runs are considered successful.

For \ac{ADAM} we here show the optimization results with $\eta=0.06$ which similarly display a deterioration in accuracy for system sizes beyond $N=26$.
It is important to note that the failed \ac{ADAM} runs are interrupted after $5\cdot 10^4$ iterations and convergence with additional runtime is not excluded in general.
The question is then: How many iterations are needed for convergence?
We observe a polynomial scaling of the required iterations in the system size up to a transition point $N^*(\eta)$ which depends on the chosen learning rate.
Above this system size \textit{both} successful and failing runs take much longer and exceed the set budget of $5\cdot 10^4$ iterations.
In fig.~\ref{fig:tfi_nondist_prec} we present the \ac{ADAM} runs for a medium learning rate in order to demonstrate the described behavior but not the best possible performance of the \ac{ADAM} optimizer. We present a more detailed analysis of the influence of the learning rate on the performance of \ac{ADAM} in appendix~\ref{sec:app_eta_Adam}.

\ac{NatGrad} shows reliable convergence to a global minimum for all sampled initial parameters.
The number of epochs to convergence scales polynomially with the system size and there is little variance in the required number of epochs. 

\begin{figure}
\flushleft
\includegraphics[width=0.48\textwidth]{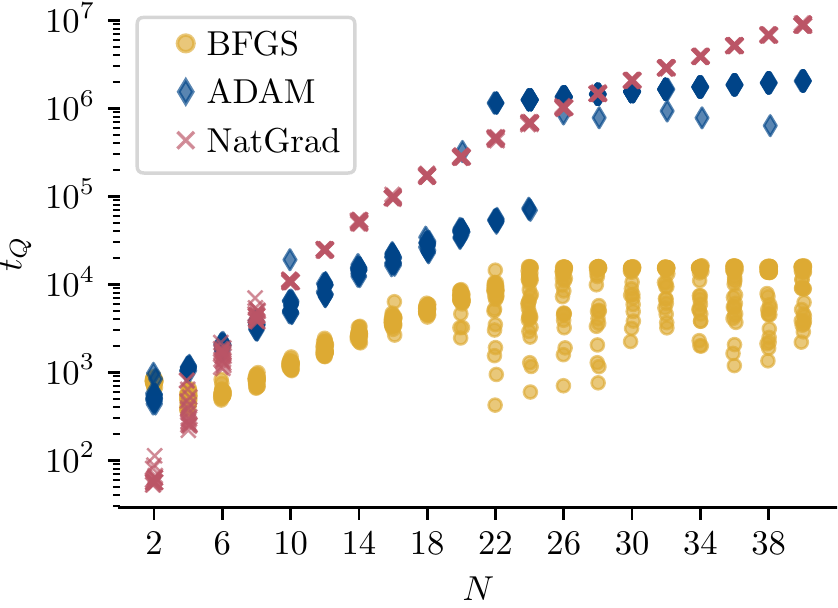}
\caption{Estimated runtimes on a quantum computer for the optimization tasks shown in fig.~\ref{fig:tfi_nondist_prec} based on the scalings in tab.~\ref{tab:optimizercost}.
We only show successful runs with $\gd_{min}\leq10^{-5}$ and note that none of the \ac{ADAM} optimizations for $N\geq 30$ attained the full precision of $10^{-10}$ such that the scaling is truncated.
\label{fig:tfi_nondist_tq}}
\end{figure}

Using the scalings discussed in more detail in sec.~\ref{sec:optcost}, taking the translation symmetry of the \ac{TFIM} into account and employing the estimates $N_M/N_a\approx 10$ \cite{McArdle_Yuan_19} and $t_3\approx t_2\approx t_1$ (see sec.~\ref{sec:optcost} for definitions of these quantities) we show the expected optimization durations on a quantum computer in fig.~\ref{fig:tfi_nondist_tq}.
Due to the increased cost per epoch and a similar scaling of the number of iterations for all optimizers, the cost for \ac{NatGrad} are considerably higher than those for \ac{BFGS} and \ac{ADAM} in the regimes in which they converge and \ac{ADAM} does not suffer from the sudden increase in required epochs.
We expect the scaling for \ac{ADAM}, which is truncated in fig.~\ref{fig:tfi_nondist_prec} due to our epoch budget, to yield quantum runtimes comparable to those of \ac{NatGrad}.
As we show in appendix \ref{sec:app_eta_Adam}, reducing the learning rate makes bigger system sizes accessible to \ac{ADAM}, but also rather drastically increases run times because of slower convergence.

The structure of the investigated Hamiltonian has a major influence on the scalings as the translation symmetries in the presented spin chain models reduce $K_H$ to a small constant leading to high relative cost of obtaining the Fubini matrix.
For chemical systems, for example, with at least quadratic scaling of $K_H$ in $N$ and depending on the ansatz class, the relative additional cost per epoch for \ac{NatGrad} can be significantly smaller, which in combination with the unreliable convergence of \ac{ADAM} from system size $N^*$ onwards would make \ac{NatGrad} an attractive optimization technique. 

In summary, we find the \ac{BFGS} optimizer to run into convergence problems already for medium sized systems, \ac{ADAM} to take a large number of epochs with a transition into unpredictable cost at a certain system size and \ac{NatGrad} to exhibit reliable convergence with fewer epochs than \ac{ADAM}, but an overall high cost when running on a real quantum computer.
Furthermore, the success of both commonly used optimizers, \ac{BFGS} and \ac{ADAM}, strongly depends on the initial parameters whereas \ac{NatGrad} shows stable convergence and a small variance of the optimization duration.

\subsection{Overparametrization by adding Y layers}
We now extend the optimal \ac{QAOA} circuit for the \ac{TFIM} by adding redundant layers of Pauli $Y$ rotations. 
These additional rotations can be deactivated by setting their variational parameter $\gk$ to zero. This means in particular that the new ansatz classes still contain the ground state and simply introduce a form of overparametrization.

As Pauli $Y$ rotations cannot be represented in the free fermion basis of the Hamiltonian (see eqn.~(\ref{eq.res:ff_mapping})), the overparametrized class can be seen as breaking a symmetry.
This means that for any given $\gk\neq 0$, the ansatz state will not be a global minimum and it will be crucial for an optimization algorithm to find the submanifold with $\gk=0$. 
This is clear for a single additional layer of gates, but we expect it to hold for multiple layers as well.
Although the present situation is artificially constructed and the broken symmetry is manifest, similar behavior is expected in systems where we do not have an analytical solution.
More generally, even for a suitable ansatz class a very specific configuration of the variational parameters is necessary to find the ground state and the chosen optimization algorithm consequentially should be resilient to local minima.
Our choice of overparametrization leads to such local minima, constructing an optimization problem that can be used as a test for the resilience of the optimizer.

\begin{figure}
\flushleft
\includegraphics[width=0.48\textwidth]{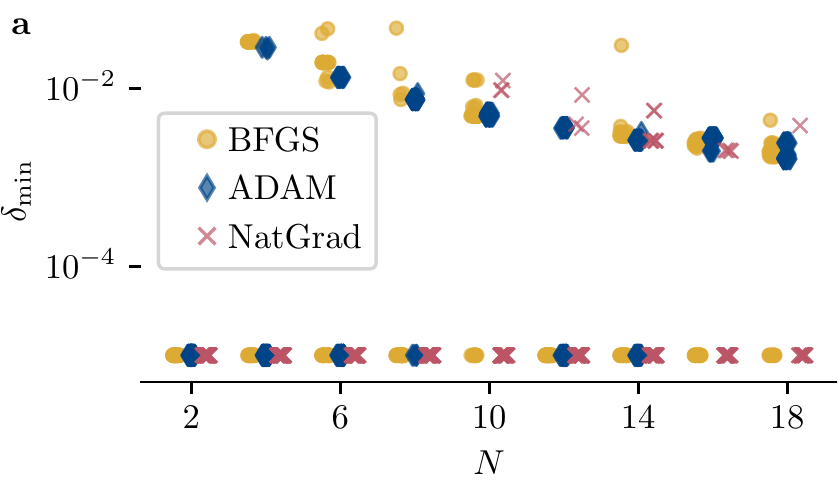}

\vspace{0.25cm}
\includegraphics[width=0.48\textwidth]{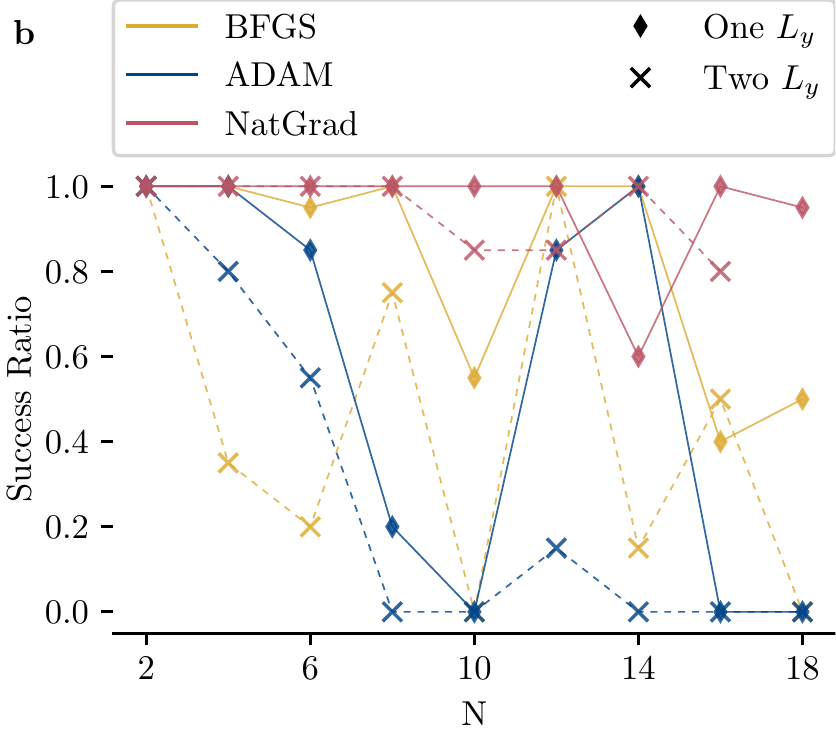}
\caption{(a) Achieved precisions $\gd_\text{min}$ and (b) fraction of successful optimizations with the three optimizers on \ac{QAOA} circuits extended by one or two Pauli $Y$-rotation layers. 
Successful optimization runs and those only converging locally are separated by a gap in the attained minimal precision and in contrast to fig.~\ref{fig:tfi_nondist_prec} the iteration budget is almost never consumed entirely.
Instead the optimization is completed -- yielding either a global or a local minimum.
\label{fig:tfi_ylay_ratio}}
\end{figure}

We look at two configurations of the extended circuits with y-rotation layers included at positions $\left\{\left\lfloor \frac{N}{4}\right\rfloor\right\}$ and $\left\{\left\lfloor \frac{N}{4}\right\rfloor,\left\lfloor \frac{N}{2}\right\rfloor-1\right\}$ respectively.
With this choice we avoid special points in the circuit and expect these setups to properly emulate the problem of (additional) local minima.

\begin{figure}
\flushleft
\includegraphics[width=0.48\textwidth]{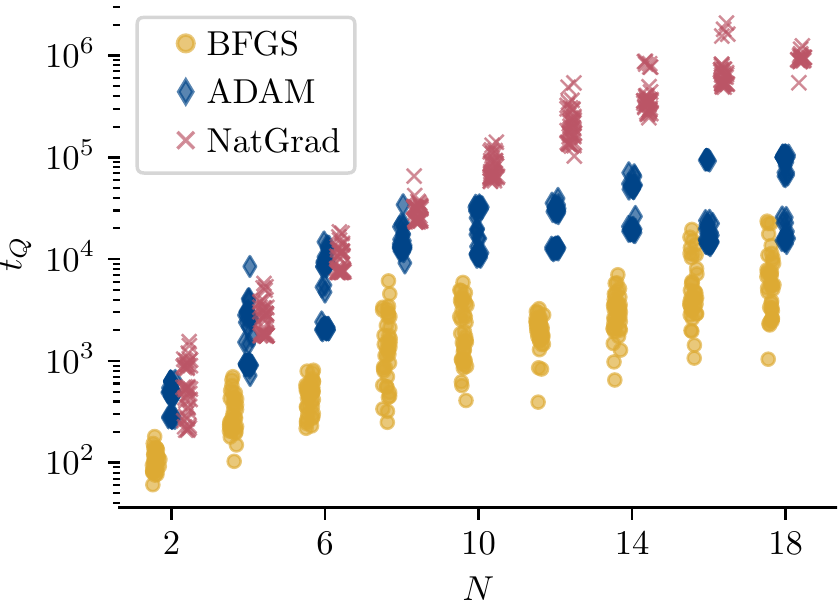}
\caption{Estimated runtime scaling on a quantum computer for the optimizations in fig.~\ref{fig:tfi_ylay_ratio} based on tab.~\ref{tab:optimizercost} and the same assumptions as in fig.~\ref{fig:tfi_nondist_tq}.
Here we also include unsuccessful instances and for the \ac{ADAM} optimizer the lower branch corresponds to successful minimizations.
\label{fig:tfi_ylay_tq}}
\end{figure}

Again we sample 20 positions in parameter space close to the origin and initialize the three optimizers at these points, resulting in the precisions and success ratios shown in fig.~\ref{fig:tfi_ylay_ratio} together with the estimated quantum computer runtimes in fig.~\ref{fig:tfi_ylay_tq}.
We observe a clear distinction between the optimizations that succeed to find a global minimum and those which converge to a local minimum only, such that we obtain a well-defined success ratio for this numerical experiment.
In contrast to the results for the minimal \ac{QAOA} circuit, no intermediate precisions caused by a finite iteration budget occur. 
All optimizers suffer from the introduced gates as they show convergence to local minima for system sizes they tackled successfully without overparametrization.

For \ac{BFGS}, this effect appears for some system sizes for one layer of Pauli $Y$ rotations but is much stronger for two additional layers, reducing the fraction of globally minimized runs to less than 50\% for multiple system sizes.
We do not claim a scaling behaviour with the system size but note an alternating pattern for the configuration with two $Y$ layers, demonstrating large fluctuations of the success ratio (\cf in particular system sizes $10$ and $
12$ for two $Y$ layers).

For the \ac{ADAM} optimizer we use a comparably small learning rate of $\eta=0.02$ which pushes the jump of the optimization duration observed before well out of the treated system size range.
Nonetheless, we observe runs stuck in local minima already for small systems without exceeding the iteration budget such that in contrast to sec.~\ref{sec:res_tfi_qaoa} allowing for a longer runtime would not improve the performance.
Also for \ac{ADAM} the fraction of successful instances fluctuates with the system size but in particular for two Pauli $Y$ rotation layers the effect becomes stronger for bigger systems and no successful runs were observed for $N\geq14$.

The performance of \ac{NatGrad} on the other hand, for which we reduced the learning rate to $\eta=0.05$, is more reliable and the success rate is the best for most of the circuits, with few exceptions.
In particular there are only few system sizes with local convergence for one and two additional degrees of freedom each and overall the success rate of \ac{NatGrad} does not drop below $60\%$.

For all optimizers we confirm that successful runs deactivate the additional Pauli $Y$ rotation layers by setting the corresponding parameters to 0 and that all optimizations with worse precision failed to do so, leading to a local minimization only.
The quantum runtimes demonstrate the expected scaling with \ac{NatGrad} as the most expensive optimizer, where the small iteration count compensates the increased cost per epoch for small systems. However, the increased effort is rewarded with significantly higher success rates, making \ac{NatGrad} a strong choice for (potentially) overparametrized \ac{VQE} optimization.
We again note that the relative cost of the Fubini matrix are high for spinchain systems and that the reduced number of epochs required by \ac{NatGrad} will have a bigger impact in other systems.

Overall our numerical experiments with the extended \ac{QAOA} circuits for the \ac{TFIM} demonstrate the fragility of the three tested optimizers to perturbations of the ansatz class.
A significant decrease in performance is caused by overparametrization outside of the symmetry sector of the model and the \ac{QAOA} ansatz class.
All algorithms were successful for the original \ac{QAOA} circuits on the considered system sizes such that the reduced success ratio can directly be attributed to the extension of the ansatz class.
This is in contrast to machine learning settings where heavy overparametrization is essential to make the cost function landscape tractable to local optimizers like \ac{ADAM}.
The strong fluctuations over the tested system sizes indicate that more repetitions of the optimization would be required to resolve systematic behaviour.

We note that the \ac{BFGS} algorithm in some instances converges to a local minimum although it has access to non-local information via its line search subroutine.
In particular in the presence of two misleading parameters in the search space, the local information determining the one-dimensional subspace does not seem to suffice any longer to find the global minimum, even though the approximated Hessian is used. 
For the \ac{ADAM} optimizer the initial gradient leads to an activation of symmetry breaking layers and due to the restriction to local information the algorithm is not able to leave the resulting sector of the search space with local minima it enters initially.
\ac{NatGrad} also is affected by the limitation to local information but because of the access to geometric properties of the ansatz state class it was on average less likely to leave the Pauli $Y$-rotation layers activated.
We attribute this to the fact that \ac{NatGrad} performs the optimization in the locally undeformed Hilbert space by extracting the influence of the parametrization. 
As a consequence the optimizer does not follow the incentive to activate the Pauli $Y$ rotations at the beginning when given the same gradient as \ac{ADAM}, but stays within the minimal parameter subspace.
A better foundation for this intuition and the observed exceptions will be subject to further investigations of \ac{NatGrad}.

In general, one could expect the cost function of \acp{VQE} to behave differently than those in common machine learning models as the parameters enter in a very non-linear manner via rotation gates.
We were able to demonstrate such a difference with \ac{ADAM}, which benefits from overparametrization in machine learning applications but suffers significantly from the additional parameters of the extended circuits.
The restriction of \ac{NISQ} devices to rather shallow circuits implies much smaller numbers of variational parameters than in machine learning such that \ac{NatGrad} can be considered a viable option for \ac{VQE} optimization.

\subsection{Results on the Heisenberg model}\label{sec:res_xxz} 

\begin{figure}
\flushleft
\vspace{0.25cm}
\includegraphics[width=0.48\textwidth]{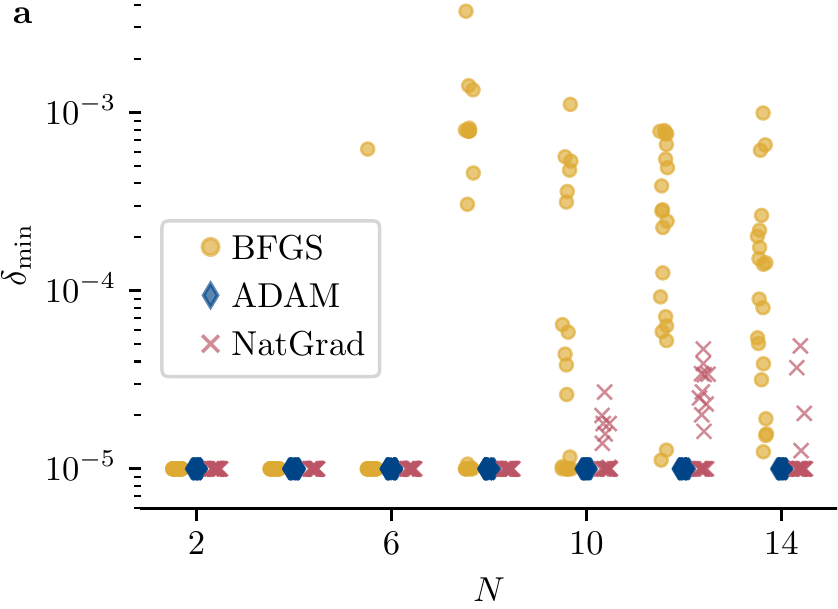}

\vspace{0.25cm}
\includegraphics[width=0.48\textwidth]{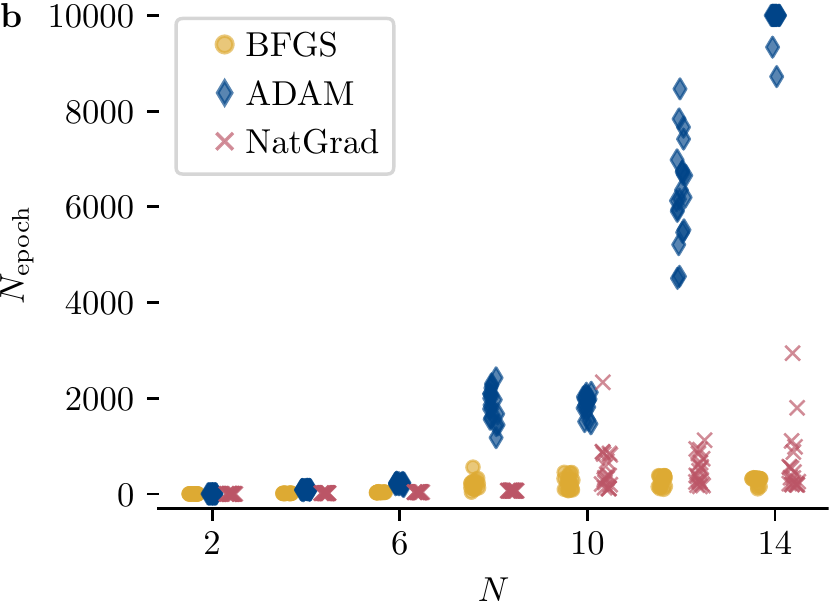}
\caption{(a) Minimal achieved precisions and (b) iteration count of the three optimizers on the ansatz in eqn.~(\ref{eq.def:xxz_ansatz}) for the \ac{XXZM} at depth $p=N$.
The circuit contains $n=3N$ parameters and the learning rates are $0.03$ and $0.1$ for \ac{ADAM} and \ac{NatGrad} respectively. The epoch count is truncated at $10000$ iterations to improve the readability.
\label{fig:xxz}}
\end{figure}

To complement the study on scaling and overparametrization in the integrable \ac{TFIM} we present here numerical results on the \ac{XXZM} with the ansatz discussed in detail in sec.~\ref{sec:xxz}.
The performance of the three optimizers, initialized at $20$ distinct points close to $0$, is shown in fig.~\ref{fig:xxz} together with the number of epochs.

The behaviour of \ac{ADAM} and \ac{BFGS} is similar to the one observed on the \ac{TFIM}, \ie \ac{ADAM} successfully achieves the target accuracy of $10^{-5}$ but shows an abrupt increase in the iteration number and \ac{BFGS} starts to fail for medium sized systems. 
The number of variational parameters at which the respective transition occurs is similar to that in the \ac{TFIM}:
The cost of \ac{ADAM} jump abruptly at $n=24$ and $n=36$ and similarly the runs with comparable learning rate for the \ac{TFIM} show (less clear) transitions at $n=26$ and $n=30$.
Likewise the \ac{BFGS} optimizer starts failing significantly at $n=24$ and $n=22$ for the \ac{XXZM} and the \ac{TFIM}, respectively.
The Hilbert space dimension however clearly differs at the transition points.
It is intuitively clear that the main influence should be due to the properties of the parameter space, but in general the physical system size might affect the performance as well by shaping the energy landscape.

The \ac{NatGrad} optimizer is less performant on the \ac{XXZM} as it is sometimes interrupted during phases of small updates.
This might indicate either convergence to a local minimum or a too small learning rate.
A preliminary further analysis showed that reducing $\eta$ in \ac{NatGrad} can prevent convergence for some instances that were optimized successfully before.
This hints to the second scenario because a reduced learning rate should generally improve the quality of \ac{NatGrad}.
This will be investigated in a follow-up study.

We note that the attained precision in failed runs does not show a consistent gap across the system sizes which makes the analysis of the performance less clear.
Nonetheless the deviation from the target precision is significantly smaller for \ac{NatGrad} than for \ac{BFGS} and if one extends the gap visible for $N=8$ and $N=10$ many instances of \ac{BFGS} are categorized as unsuccessful.

We interpret the results on the \ac{XXZM} as follows:
Some difficulties of the commonly used \ac{BFGS} optimizer and \ac{ADAM} appear also in this model already for moderate system sizes.
The size of the parameter space seems to primarily determine whether performance (\ac{BFGS}) or runtime (\ac{ADAM}) issues arise, not so much the Hilbert space dimension of the underlying many-body model.
The very reliable performance of \ac{NatGrad} seen in the \ac{TFIM} can not necessarily be generalized to other spin chain models, let alone to other classes of Hamiltonians.
However, the characteristics of the failed runs let us hope that systematic improvements to \ac{NatGrad} might be possible. 

\section{Methods}\label{sec:methods}

\subsection{Variational Quantum Eigensolver}
The framework of our work is the \ac{VQE}, a proposal to use parametrized circuits on a quantum computer in combination with classical optimization routines to prepare the ground state of a target Hamiltonian $H$.
In the first part of a \ac{VQE} one constructs a quantum circuit that contains parametrized gates.
Given input parameters $\gT$ for the circuit, a quantum computer can then prepare the corresponding ansatz state and measure an objective function, chosen to be the energy of the Hamiltonian
\begin{equation}\label{eq.def:energy_cost}
E(\gT)\coloneqq \bra{\psi(\gT)}H\ket{\psi(\gT)}
\end{equation}
and for benchmark problems with known ground state energy $E_0$, the relative error $\gd$ can be calculated as
\begin{equation}
\gd(\gT)\coloneqq\frac{E(\gT)-E_0}{|E_0|}.
\end{equation}
Additionally one can prepare modified versions of the circuit to determine auxiliary quantities like the energy gradient in the parameter space \cite{Schuld_Killoran_19}.
The second part of the \ac{VQE} scheme is an optimization strategy on a classical computer which is granted access to the quantum black box just constructed.
In the most straightforward scenario this is a black box minimization scheme, but using auxiliary quantities,  more sophisticated optimization methods can be realized as well.

There are two main theoretical challenges for successfully applying \ac{VQE}:
First, the construction of a sufficiently complex, but not overly expensive, circuit that gives rise to an ansatz class containing the ground state -- \textit{expressivity}.
Second, the choice of a suitable optimizer that is able to search for the ground state within the created parameter space -- \textit{efficiency}.
The two challenges are often seen as independent, but explicit algorithms using information gathered about the variational space during optimization phases for adjusting the ansatz have been proposed as well, some of which are inspired by concrete applications in quantum chemistry or by evolutionary  strategies \cite{Grimsley_Mayhall_19, Tang_Economou_19, Ostaszewski_Benedetti_19, Rattew_Wood_19}.

We now establish some notation for the general \ac{VQE} setting where we assume the  most common objective: Finding the ground state energy of a Hamiltonian $H$.
Starting from an initial product state $\ket{\bar{\psi}}$, we apply parametrized unitaries $\{U_j(\gT_j)\}_{1\leq j\leq n}$ to construct the ansatz state
\begin{equation}
\ket{\psi(\gT)}\coloneqq \p{j}{N}{1} U_j(\gT_j)\ket{\bar{\psi}}.
\end{equation}
The parameters are typically initialized randomly close to zero to avoid the barren plateau problem \cite{McClean_Neven_18}.
For this work, the unitaries are going to be translationally invariant layers of one- or two-qubit rotations; Consider for instance 
\begin{align}
  L_{zz}(\gT_j) &\coloneqq \p{k}{1}{N}\exp\left[-\frac{i\gT_j}{2} Z^{(k)}Z^{(k+1)}\right]\\
  &=\exp\left[-\frac{i\gT_j}{2} \s{k}{1}{N}Z^{(k)}Z^{(k+1)}\right] ,
\end{align}
where we identified the qubits with index 1 and N+1, \ie we adopt periodic boundary conditions.
The ordering of the gates within a layer is not relevant because they commute but for convenience we write them such that terms acting on the first qubits are applied first. 
$Z^{(k)}$ is the Pauli $Z$ operator acting on the $k$-th qubit and we tacitly assume the tensor product between operators that act on distinct qubits as well as the missing tensor factors of identities. 
Compared to proposed ansatz circuits that employ full Hamiltonian time evolution $\exp[-i\gT H]$ (see sec.~\ref{sec:qaoa}), such a layer is rather easily implemented on present quantum machines because it only requires linear connectivity and one type of two-qubit rotation.
There have been many proposed circuits to generate ansatz classes for a variety of problems, all of which can be boiled down to combining rotational gates and possibly other fixed gates such as the \acs{CNOT} or \acs{SWAP} gate (see sec.~\ref{sec:ansatze}).
For the presented optimization methods the derivatives \wrt the variational parameters $\{\gT_j\}_j$ are important and for the above example we observe the special structure of translationally symmetric layers of Pauli rotation gates:
\begin{equation}\label{eq.vqe:layer_derivative}
\frac{\partial}{\partial\gT_j}L_{zz}(\gT_j) = \left(-\frac{i}{2}\s{k}{1}{N}Z^{(k)}Z^{(k+1)}\right) L_{zz}(\gT_j) .
\end{equation}
The derivative only produces an operator prefactor, and all prefactors can be summarized because the single gates commute. 
While the basic gates composing a unitary $U_j(\gT_j)$ typically take the form of (local) Pauli rotations, the full unitary often is more complex than the above layer and in particular the terms in $U_j$ do not need to commute. 
However, the structure of rotations enables us in general to evaluate required expressions involving derivatives on a quantum computer, either via measurements of rotation generators or via ancilla qubit schemes.

\subsubsection{A selection of ansatz classes}\label{sec:ansatze}
Among the ansatz families proposed in the literature we present the following which are used frequently and of which two are directly connected to this work:
\paragraph{QAOA}\label{sec:qaoa}
The Quantum Approximate Optimization Algorithm was first proposed by Farhi, Goldstone and Gutmann \cite{Farhi_Goldstone_14} in 2014 for approximate solutions to (classical) optimization problems by mapping them to a spinchain Hamiltonian.
The algorithm looks similar to adiabatic time evolution methods with an inhomogeneous time resolution which is rather coarse for typical circuit depths.
A lot of work has been put into proving properties of the \ac{QAOA} both, in general and for certain problem types, including extensions to quantum cost Hamiltonians \cite{Morales_Biamonte_19, Lloyd_18, Hastings_19, Farhi_Harrow_16}.
A the same time the algorithm has been refined, extended, and characterized on the basis of heuristics and numerical experiments, gaining insight into its properties beyond rigorous statements \cite{Wang_Rieffel_18, Mbeng_Santoro_19, Ho_Hsieh_19, Niu_Chuang_19, Akshay_Biamonte_19}.

The \ac{QAOA} circuit is constructed as follows: 
For a \textit{cost Hamiltonian} $H_S$ and a so-called \textit{mixing Hamiltonian} $H_B$ one alternatingly applies the unitaries $\exp\left[-i\gt_j H_S\right]$ and $\exp\left[-i\gf_j H_B\right]$ $p$ times, giving rise to a \ac{VQE} ansatz class with 'time' parameters $\{\gt_j, \gf_j\}_{1\leq j\leq p}$. 
Originally, the system Hamiltonian would encode a classical optimization problem and thus be diagonal while the mixing Hamiltonian was chosen to be off-diagonal and specifically has been kept fixed to the original $H_B=\s{k}{1}{N} X^{(k)}$ for many investigations. 
However, new choices of mixers have been proposed and investigated as well, giving rise to the more general \ac{QAOa}\cite{Hadfield_Rieffel_19, Wang_Rieffel_19, Akshay_Biamonte_19}.

Note that for quantum systems, the terms comprising the Hamiltonian $H_S$ do not commute in general such that very large gate sequences would be necessary to realize the exact \ac{QAOA} approach including $\exp\left[-i\gt H_S\right]$. 
In practice these blocks commonly are broken up in a Trotter-like fashion instead, yielding circuits that are implemented more readily but deviating from the original ansatz. 
For the \ac{TFIM}, such a modified \ac{QAOA} ansatz has been studied intensively \cite{Wang_Rieffel_18, Ho_Hsieh_19, Mbeng_Santoro_19} and we are going to use it as a starting point for our investigations.

\paragraph{Adaptive ans\"atze}
Most prominently for this type of ans\"atze, \acs{ADAPT-VQE} tackles both the construction of a suitable ansatz class and the optimization within the constructed parameter space.

Instead of a fixed ansatz circuit layout, \acs{ADAPT-VQE} takes a pool of gates as input and iterates the two steps of the \ac{VQE} scheme:
After rating all gates the most promising one is appended to the circuit (construction) and afterwards all the circuit parameters are optimized (minimization).
The optimized parameters from the previous step are then used for both, the rating of the gates for the next construction step and the initialization for the following optimization, where newly added gates are initialized close to the identity.
For both, the concept of allowed gates and the gate rating criteria, there are multiple options and we refer the reader to \cite{Grimsley_Mayhall_19, Tang_Economou_19} for more detailed descriptions.

Besides \acs{ADAPT-VQE}, multiple other methods which grow the ansatz circuit in interplay with the optimization have been proposed and demonstrated, including \textit{Rotoselect} \cite{Ostaszewski_Benedetti_19} and \acs{EVQE} \cite{Rattew_Wood_19}.
These demonstrations include the solution of 5-qubit spinchains and small molecules (lithium hydride, beryllium dihydride and a Hydrogen chain) to chemical precision using simulations with and without sampling noise or quantum hardware.

We will not be using any adaptive scheme in our work, but our results on stability and overparametrization raise serious doubts as to the reliability of any adaptive ansatz method.

\subsection{Optimizers}
A variety of optimizers have been used in the context of variational quantum algorithms.
These optimizers are inspired by classical machine learning and can be sorted according to the order of information required about the cost function.
Zeroth-order or direct optimization methods only evaluate the function itself, first-order methods need access to the gradient, and second-order optimization need access to the Hessian of the cost function, or some other metric reflecting the local curvature of the learning landscape.

\subsubsection{Direct optimization}
The most naive approach to optimizing a function over an input space is to simply ``look at all possible inputs'', \ie to set up a grid and to evaluate the function on all vertices of the grid. 
Even though it is unlikely to find the minimum in this manner directly, subsequent refinements of the grid around potential minima make global optimization possible. 
On the one hand this method becomes exponentially expensive in the number of parameters and a 15-dimensional grid generated by only two values per parameter already requires $2^{15}>3\cdot 10^4$ function evaluations. 
On the other hand, the naive grid search can be improved significantly which allows for global optimization.
This approach has been demonstrated successfully for between 15 to 20 parameter with the \acs{DIRECT} method and a budget of $2\cdot 10^5$ evaluations \cite{Kokail_Zoller_19}. 
For high-dimensional applications, \ie circuits for realistic systems with parameter count at least linear in the size of the system, any global optimization strategy seems likely to suffer from the sparse information access and to become incapable of exploring a sufficiently big fraction of the search space.

As is the case for most of the work on \acp{VQE} we will not use any direct minimization methods, supported by the estimate that those strategies become unfeasible for relevant problem sizes and demonstrated deficiencies in comparison to gradient-based techniques \cite{Romero_Aspuru-Guzik_17}.

\subsubsection{First-Order Gradient Descent}
Optimization techniques using the gradient of the cost function are at this point the most widely used in machine learning.
Starting from the simple Gradient Descent method that updates the parameters according to the gradient and a fixed learning rate, a whole family of minimization strategies has been developed.
The improved routines are inspired by physical processes like momentum, based on heuristics like adaptive learning rate schedules, or a smart processing of the gradient information as in the Nesterov Accelerated Gradient.
A review of this development can be found \eg in \cite{Ruder_16}, here we just present the first-order method we are going to use, the \ac{ADAM} optimizer.

\ac{ADAM}, which was proposed in 2014 \cite{Kingma_Ba_14}, is probably the most prevalent optimization strategy for deep feed-forward neural networks \cite{Karpathy_17} and has been used in \ac{VQE} settings as well \cite{Ostaszewski_Benedetti_19}. 
For completeness, we briefly outline the \ac{ADAM} optimizer:
Given the cost function $E(\gT)$, where $\gT$ recollects all variational parameters, a starting point $\gT^{(0)}$ and a learning rate $\eta$, Gradient Descent computes the gradient $\nabla E(\gT^{(t)})$ at the current position and accordingly updates the parameters rescaled by $\eta$:
\begin{equation}
\gT^{(t+1)} = \gT^{(t)}-\eta \nabla E(\gT^{(t)}) .
\end{equation}
As the gradient points in the direction of steepest ascend, the parameter update is directed towards the steepest descend of the cost function and for $\eta$ small enough, the convergence towards a minimum can be understood intuitively. 
Small learning rates yield slow convergence which increases the cost of the optimization whereas choosing $\eta$ too large leads to overshooting and oscillations which might prevent convergence. 
Furthermore, although the optimizer will diagnose convergence to a minimum due to a vanishing gradient, it cannot distinguish between local and global minima.

In order to fix both issues, \ie the need for an optimally scheduled learning rate and the liability of getting stuck in local minima, various improvements have been proposed and \ac{ADAM} uses several of these upgrades. 
The first feature is an \textit{adaptive, componentwise learning rate}, which was introduced in \acs{AdaGrad} \cite{Duchi_Hazan_11} and improved in \acs{RMSprop} \cite{Hinton_12} to avoid suppressed learning.
The second feature \ac{ADAM} uses is \textit{momentum}, which is inspired by the physical momentum of a ball in a landscape with friction. 
This is realized by reusing past parameter upgrades weighted with an exponential decay towards the past and enables \ac{ADAM} to overcome some local minima.
The final form of the \ac{ADAM} algorithm is as follows:
Initialize with hyperparameters $\{\eta,\gb_1,\gb_2,\ge\}$, momentum $m^{(0)}=0$, average squared gradient $v^{(0)}=0$ and initial position $\gT^{(0)}$.
At the t-th step, compute the gradient and update the momentum and the cumulated squared gradient as
\begin{align}
m^{(t)} &= \frac{\gb_1-\gb_1^t}{1-\gb_1^t}m^{(t-1)}+\frac{1-\gb_1}{1-\gb_1^t}\nabla E(\gT^{(t)}) ,\\
v^{(t)} &= \frac{\gb_2-\gb_2^t}{1-\gb_2^t}v^{(t-1)}+\frac{1-\gb_2}{1-\gb_2^t}\left(\nabla E(\gT^{(t)})\right)^{\odot 2}
\end{align}
where $x^{\odot 2}$ denotes the elementwise square of a vector $x$.
The parameter update then is computed from these updated quantities via
\begin{equation}
\gT^{(t+1)}=\gT^{(t)}-\frac{\eta}{\sqrt[{\odot}]{v^{(t)}}+\ge}m^{(t)}
\end{equation}
with the square root of $v^{(t)}$ taken elementwise. 
Besides the learning rate $\eta$, we identify the hyperparameters $\gb_1$ and $\gb_2$ as exponential memory decay factors of $m$ and $v$ respectively and the small constant $\ge$ as regularizer, which avoids unreasonably large updates in flat regions and division by zero at initialization or for irrelevant parameters.

Because of the advanced features that \ac{ADAM} uses, it has been very successful at many tasks and even though there are applications for which more basic gradient-based optimizers can be advantageous, we choose \ac{ADAM} to represent the family of local first-order optimizers.

\subsubsection{BFGS optimizer} 
The second optimizer we look at is the \ac{BFGS} algorithm, which was proposed by its four authors independently in 1970 \cite{Broyden_70, Fletcher_70, Goldfarb_70,Shanno_70} . 
Using first-order resources only it approximates the Hessian of the cost function and performs global line searches in the direction of the gradient transformed by the Hessian inverse.
Therefore it is a global quasi second-order method using local first-order information and its categorization is not obvious.
The algorithm is initialized with the starting point $\gT^{(0)}$ and a first guess for the approximate Hessian $H^{(0)}$ of the cost function $E$, which usually is set to the identity.
At each step of the optimization one determines the gradient, computes the direction 
\begin{equation}
n^{(t)} = {H^{(t)}}^{-1}\nabla E(\gT^{(t)})
\end{equation}
and performs a line search on $\{\gT^{(t)}+\eta\; n^{(t)}|\eta\in\mathbb{R}\}$ which yields the optimal update in that direction and can optionally be restricted to a  bounded parameter subspace.
Given the new point in parameter space, $\gT^{(t+1)}$, the change in the gradient $D^{(t)}=\nabla E(\gT^{(t+1)})-\nabla E(\gT^{(t)})$ is calculated and used to update the approximate Hessian via
\begin{equation*}
H^{(t+1)}=H^{(t)}+\frac{D^{(t)}{D^{(t)}}^T}{\eta^{(t)} {D^{(t)}}^T n^{(t)}}-\frac{H^{(t)}n^{(t)}{n^{(t)}}^TH^{(t)}}{{n^{(t)}}^TH^{(t)}n^{(t)}} .
\end{equation*}
As the parameter updates are found via line searches, the \ac{BFGS} algorithm is not strictly local but due to its use of local higher-order information, the global search is much more efficient than direct optimization. 
The method has been successful in many applications and currently is of widespread use for \acp{VQE}. \cite{Guerreschi_Smelyanskiy_17, Mbeng_Santoro_19, Wang_Rieffel_19, Grimsley_Mayhall_19, Romero_Aspuru-Guzik_17, Gard_Barnes_20}

\subsubsection{Natural Gradient Descent}
The third optimization strategy we use is the \ac{NatGrad} \cite{Stokes_Carleo_19, Amari_98, Harrow_Napp_19}, which due to its increased cost per epoch is not adopted very often in machine learning settings itself but is connected to some successful methods.
As an example, Stochastic Reconfiguration which is closely related to \ac{NatGrad}\cite{Becca_Sorella_17} recently has been shown to work well for training \acp{RBM} to describe groundstates of spin models \cite{Carleo_Troyer_17}.
Despite this success, the insights into why and under which conditions the method works remain limited and recent work has been put into understanding the learning process for the mentioned application of \acp{RBM} and the Natural Gradient Descent \cite{Park_Kastoryano_19}.
Before discussing \ac{NatGrad} and its role in the \ac{VQE} setting, we outline its update rule:
Given a starting point $\gT^{(0)}$ and a learning rate $\eta$, a step is performed by first constructing the Fubini-Study metric of the ansatz class
\begin{equation}\label{eq.natgrad:fubini}
  \begin{split}
    \left(F_t\right)_{ij} &\coloneqq \real{\ip{\partial_{i}\psi^{(t)}}{\partial_{j}\psi^{(t)}}}\\
    &-\ip{\partial_{i}\psi^{(t)}}{\psi^{(t)}}\ip{\psi^{(t)}}{\partial_{j}\psi^{(t)}}
  \end{split}
\end{equation}
at the current position and then updating the parameters via
\begin{equation}
\gT^{(t+1)} = \gT^{(t)}-\eta\;{F^{(t)}}^{-1}\nabla E(\gT^{(t)})
\end{equation}

where we abbreviated $\ket{\psi^{(t)})}\coloneqq\ket{\psi(\gT^{(t)})}$ and $\ket{\partial_i\psi^{(t)}}\coloneqq\frac{\partial}{\partial\gT_i}\ket{\psi(\gT^{(t)})}$.

The Fubini-Study metric is the quantum analogue of the Fisher information matrix in the classical Natural Gradient \cite{Amari_98}. 
It describes the curvature of the ansatz class rather than the learning landscape, but often performs just as well as Hessian based methods.
In order to avoid unreasonably large updates caused by very small eigenvalues of $F$ in standard Natural Gradient Descent $\eta$ has to be chosen very small for an unpredictable number of initial learning steps.
Alternatively one can use \textit{Tikhonov} regularization which amounts to adding a small constant to the diagonal of $F$ before inverting it.

Even though \ac{NatGrad} is simple from an operational viewpoint, it is epochwise the most expensive optimizer of the three presented here (also see sec.~\ref{sec:optcost}).
This is due to the fact that it not only uses the gradient but, in order to construct the
(Hermitian) matrix $F$ for $n$ parameters, one also needs to evaluate $\frac{1}{2}(n^2+3n)$ pairwise overlaps of the set $\{\ket{\psi}, \ket{\partial_1\psi},\dots,\ket{\partial_n\psi}\}$ (all but $\ip{\psi}{\psi}=1$).
Depending on the gates in the ansatz circuit, each of these overlaps requires at least one and possibly many individual circuit evaluations.
For circuits containing $\tilde{n}$ simple one- or two-qubit Pauli rotation gates, the number of circuits required is $\frac{1}{2}(\tilde{n}^2+3\tilde{n})$, independent of the number of shared parameters.
Symmetries of the circuit may reduce the number of distinct terms in which case fewer quantum machine runs suffice.
  
Taking the $j$-th parametrized unitary to have $K_j$ Hermitian generators $P_{j,k_j}$, \eg Pauli words up to prefactors $\{c_{j,k_j}\}$, the factors in the second expression of $F$ take the shape of an expectation value (see also eqn.~(\ref{eq.vqe:layer_derivative}))
\begin{equation}
\ip{\psi}{\partial_j\psi}=\bra{\bar{\psi}}\p{l}{j-1}{1}U_l\dag \left[\s{k_j}{1}{K_j} c_{j,k_j}P_{j,k_j} \right]\p{l}{1}{j-1}U_l\ket{\bar{\psi}} .
\end{equation}
The first term in eqn.~(\ref{eq.natgrad:fubini}) requires slightly more complex circuits using one ancilla qubit and a depth which depends on the indices of the matrix entry \cite{Guerreschi_Smelyanskiy_17, Li_Benjamin_17, Romero_Aspuru-Guzik_17, Dallaire-Demers_Aspuru-Guzik_19}.
Both for simulation work and for applications on real quantum machines, the construction of the Fubini matrix is expected to take much more time than inverting it -- in sharp contrast to typical classical machine learning problems.
Given the scaling of the number of required circuits above and the fact that for a fixed number of qubits the depth has to grow at least linearly with the number of parameters, an asymptotic scaling of $\order{\tilde{n}^3}$ is a lower bound for the construction of the full matrix.
Standard matrix inversion algorithms do not only show smaller or equal scaling but also exhibit as prefactor the time cost of a \acs{FLOP} whereas the evaluation scaling has prefactors based on sampling for expectation values.

As the number of parameters in a typical \ac{VQE} circuit is considerably smaller than in neural networks and the circuit chosen in this work exhibits beneficial symmetries, the high cost of the method are expected to be less problematic for our setting and bearable for \ac{VQE} applications.
Indeed there have been some demonstrations of the Natural Gradient Descent and the Imaginary Time Evolution for small \ac{VQE} instances \cite{McArdle_Yuan_19, Stokes_Carleo_19, Koczor_Benjamin_19} as well as comparisons to standard gradient descent methods and imaginary time evolution for one- and two-qubit systems \cite{Yamamoto_19}.
Inspired by the classical machine learning context and aiming for reduced cost, modifications of Natural Gradient Descent have been proposed such as a (block) diagonal approximation to the Fubini-Study matrix \cite{Stokes_Carleo_19}.
We will later show that such simplifications have to be performed with caution and can disturb the optimization.

\begin{table*}[t!]
\begin{tabular}{lcl}
Operation  & Quantum cost &  \\\hline
Energy evaluation & $N_MK_Ht_1$ & Depending on measurement bases\\
Analytic gradient  & $(Kn)N_MK_Ht_1$ & Ancilla qubit required\\
Numeric gradient (sym.)  & $2nN_MK_Ht_1$ & \\
Numeric gradient (asym.) & $(n+1)N_MK_Ht_1$ & \\
\acs{SPSA} gradient &$2N_MK_Ht_1$ & \\
Fubini matrix &$\quad (Kn)^2N_at_3+(Kn)N_at_2\quad$&Ancilla qubit required\\\hline
\ac{BFGS} & $C_{\text{grad}}+\ggg N_MK_Ht_1$ & $\ggg=\order{n^{0\leq y<1}}$ expected\\
\ac{ADAM} & $C_{\text{grad}}$ & Monitoring adds $N_MK_Ht_1$ for some gradients \\
\ac{NatGrad} & $C_{\text{grad}}+C_{\text{Fubini}}$ & Cost for inverting $F$ can be neglected \\
\end{tabular}
\caption{\label{tab:optimizercost}Cost on a quantum computer for selected \ac{VQE} optimization methods and their subroutines. The optimizer cost are given per epoch, enabling us to compare the techniques beyond their simulation times with deviating scaling. We neglected terms which are comparably small for $d,n\gg1$.}
\end{table*}

\subsubsection{Optimization cost}\label{sec:optcost}

To make a fair comparison between the optimization schemes, we briefly lay out the scaling of the required operations and the resulting cost per epoch.

We will use the following notation during the comparison:
There are $n$ variational parameters in the circuit, $K_H$ terms in the Hamiltonian and on average $K=\s{j}{1}{n}K_j/n$ Pauli generators per variational parameter, with an average of $N_M$ samples required for each expectation value. 
In practice, one of course would measure whole sets of operators both from the Hamiltonian and from the Pauli generator set simultaneously, such that $K$ and $K_H$ essentially are numbers of bases in which measurements are required.
For entries of the Fubini matrix we assume $N_a$ samples for sufficiently precise measurements, which has been shown to be smaller than $N_M$ numerically; For further discussion see \cite{McArdle_Yuan_19}.
Finally, we introduce the time scale $t_x \coloneqq \frac{d}{x}t_{\text{gate}}+t_{\text{wrap}}$ where $t_1$ is needed to initialize and measure the quantum register ($t_\text{wrap}$) and perform the circuit with depth $d$ inbetween ($dt_\text{gate}$).

Evaluating the gradient of the energy function can be done with different methods yielding a tradeoff between precision and cost.
On one hand, the analytic gradient can be evaluated up to measurement precision at the expense of an ancilla qubit and a scaling prefactor $Kn$.
On the other hand there is the finite difference method, which can be performed symmetrically, asymmetrically or via \ac{SPSA}, with cost prefactors $2n$, $n+1$ and $2$, respectively.
This means that robustness to imprecise gradients in general is a relevant property of any optimization scheme used for \acp{VQE} because these gradients are much cheaper to evaluate.
Computing the Fubini-Study metric requires two terms and although the measurement cost scales with $\order{(Kn)^2}$ for the first and with $\order{Kn}$ for the second, we keep both terms in the overall cost scaling because the \ac{VQE} regime implies moderate values of $Kn$.

For the scalings presented in table \ref{tab:optimizercost} we assume a homogeneous distribution of the variational gates in the circuit and that similar numbers of samples $N_M$ are required to measure expectation values of the Hamiltonian terms within one basis and each derivative for all gradient methods.

For the full optimization algorithms the cost are given per epoch as we do not have access to generic scaling of epochs to convergence.
Using the cost per epoch one can rescale the optimization cost from epochs to estimated runtime on a quantum computer beyond estimates that are based on the classical simulation runtimes.
For the \ac{BFGS} algorithm we can not predict the number $\ggg$ of evaluations that are required for the line searches but our numeric experiments and the linear scaling of the cost for non-\ac{SPSA} gradients suggest that it can be neglected as compared to the gradient computation.

For the quantum runtime scalings shown in figs. \ref{fig:tfi_nondist_tq} and \ref{fig:tfi_ylay_tq} we give the time in units of $t_{eval}=N_MK_Ht_1$, assumed $N_M/N_a\approx 10$ and approximated $t_1\approx t_2\approx t_3$.
\subsection{Models}\label{sec:models} 
\subsubsection{\acl{TFIM}}\label{sec:tfi} 
Our main model is the \ac{TFIM} on a spinchain with \ac{PBC}. 
Its Hamiltonian reads
\begin{equation}\label{eq.def:H_TFI}
H_\text{TFI}= H_S + H_B \coloneqq - \s{k}{1}{N} Z^{(k)} Z^{(k+1)} - t \s{k}{1}{N} X^{(k)}
\end{equation}
where we identify the sites $1$ and $N+1$ because of the \ac{PBC} and $t$ is the transverse field. 
For $t=0$, the system is the classical Ising chain, which is also called ring of disagrees and is a special case of the \textit{MaxCut} problem \cite{Farhi_Goldstone_14, Wang_Rieffel_18}. 
For $t\neq 1$ the problem is no longer motivated by a classical optimization task and for the critical point $t=1$ the ground state exhibits long-ranged correlations.

The ground state of the \ac{TFIM} is found analytically by mapping it to a system of non-interacting fermions, such that the transformed Hamiltonian can be diagonalized \cite{Lieb_Mattis_61}. 
The translational invariance of the Hamiltonian is crucial for this step and it will be important that only a small number of different (Pauli) terms can be mapped to \textit{non-interacting} fermions simultaneously.
We show the explicit computation via the Jordan-Wigner transformation in appendix \ref{sec:app_tfi}, it can also be found in \eg \cite{Wang_Rieffel_18}.
Here we summarize the action of the mapping on the terms in the Hamiltonian which also generate the \ac{QAOA} circuit (see eqn.~(\ref{eq.res:tfi_alphas}) for the definition of $\ga_q$): 
\begin{align}\label{eq.res:ff_mapping}
  \s{k}{1}{N} Z^{(k)}Z^{(k+1)} & \longrightarrow\\
  (N-2r)+2&\bigoplus_{q=1}^{r} [\cos\ga_q \;Z +\sin\ga_q\;Y], \NN\\ 
  \s{k}{1}{N} X^{(k)} & \longrightarrow (N-2r)+2\bigoplus_{q=1}^{r} Z
\end{align}
where the expressions on the right are understood in a \textit{fermionic operator basis}. 
The ground state of $H_\text{TFI}$ is just the product of the single-fermion ground states in momentum basis and we can write out the state and its energy as
\begin{align}
  E_{0} &=-E'-2\s{q}{1}{r} \sqrt{1+t^2+2t\cos \ga_q}\;, \quad \w/\\
  \ga_q &\coloneqq \cas{(2q-1)\pi/N}{N=2r}{2q\pi/N}{N=2r+1}\label{eq.res:tfi_alphas}\\
  E'  &\coloneqq \cas{0}{N=2r}{1+h}{N=2r+1}.
\end{align}

Because of the free fermion mapping, we can not only obtain the exact ground state of the system but also justify the success of the modified \ac{QAOA} circuit for the \ac{TFIM}.
As mentioned in sec.~\ref{sec:qaoa}, the original \ac{QAOA} proposal would use the system Hamiltonian and a mixing term as generators for the parametrized gates. 
For the \ac{TFIM}, however, separating the nearest-neighbour interaction terms $H_S$ from the transverse field terms $H_B$ recombines the latter with the mixing unitary next to it absorbing one variational parameter per block.
The resulting parametrized circuit contains two types of translationally invariant layers, $L_x(\gf)$ and $L_{zz}(\gt)$, of one- and two-qubit rotation gates, respectively.
Starting in the ground state of $H_B$, that is $\ket{\bar{\psi}}=\ket{+}^{\otimes N}$, we alternatingly apply these two layers $p$ times starting with $L_{zz}$.
In the free fermion picture this translates to $\ket{\bar{\psi}}=\ket{0}^{\otimes r}$ and to rotations of the $r$ fermionic states about the z-axis ($L_x$) and an axis $e_q=(0,\sin\ga_q, \cos\ga_q)^T$ which depends on the fermion momentum $q$ ($L_{zz}$).

For $t=0$ one can prove that this circuit can prepare the ground state exactly if and only if $p\geq r$ \cite{Mbeng_Santoro_19}, whereas for the case $t\neq 0$ only numerical evidence and a non-rigorous explanation support this claim \cite{Ho_Hsieh_19}.
This explanation compares the number of independent parameters, $2p$ to the number of constraints from fixing the state of $r$ free fermions, $2r$.
While solvability would be implied for a linear system, the given problem is non-linear and the argument remains on a non-rigorous level.

Finally, the equivalence to a system of free fermions has a practical implication for our simulations of the \ac{QAOA} circuit:
Storing the state of $r$ free fermions just requires memory for $2r$ complex numbers.
Applying the entire circuit needs $2pr$ two-dimensional matrix-vector multiplications, which is contrasted by $2pN$ matrix-vector multiplications in $2^N$ dimensions for a full circuit simulation in the qubit picture.
Using the fermionic basis for numerical simulations, results on the \ac{VQE} optimization problem for up to $N=200$ and $p>120$ have been obtained for $t=0$ \cite{Mbeng_Santoro_19}.

\subsubsection{\acl{XXZM}}\label{sec:xxz}
As a second model we consider the 1D \ac{XXZM} with \ac{PBC} which is defined by
\begin{equation}
H_\text{XXZ}=\s{k}{1}{N} \left[X^{(k)}X^{(k+1)}+Y^{(k)}Y^{(k+1)}+\gD Z^{(k)}Z^{(k+1)}\right] .
\end{equation}
$\gD$ is the anisotropy parameter. As in the \ac{TFIM}, the sites $1$ and $N+1$ are identified.
The Bethe ansatz reduces the eigen value problem for the \ac{XXZM} to a system of $N/2$ non-linear equations that can be solved numerically with an iterative scheme \cite{Karbach_Tobochnik_97, Karbach_Mueller_98}. 
This results in polynomial cost for computing the ground state energy but does not yield a simple ansatz class to construct the ground state on a quantum computer or a simulation scheme of reduced complexity.

We therefore use the \ac{XXZM} as a second benchmark which models the application case more closely:
We do not know a finite gate sequence that contains the ground state but instead employ circuits composed of symmetry-preserving layers which we found to be relatively successful in experiments. 
The ansatz we choose is the first-order Trotterized version of the unitary time evolution with the system Hamiltonian applied to a antiferromagnetic ground state:
\begin{align}\label{eq.def:xxz_ansatz}
\ket{\psi(\gT)}&=\p{j}{L}{1} L_{zz}(\gt_j)L_{yy}(\gk_j)L_{xx}(\gf_j)\ket{\bar{\psi}} \\ \ket{\bar{\psi}}&=\frac{1}{\sqrt{2}}\left(\ket{01}^{\otimes N/2}\pm\ket{10}^{\otimes N/2}\right)
\end{align}
where we only treat even $N$ and $\ket{\bar{\psi}}$ is chosen symmetric under translation for $(N\mod 4)=0$ and antisymmetric for $(N\mod 4)=2$ in anticipation of the exact solution via the Bethe ansatz.
We found this circuit to be more successful at finding the ground state than the \ac{QAOA} circuit.
Even though the terms $\s{k}{1}{N} X^{(k)}X^{(k+1)}$ and $\s{k}{1}{N} Y^{(k)}Y^{(k+1)}$ do not preserve the magnetization in the $Z$-basis in general they do so within the sector described by the above ansatz.

\subsection{Simulation Details}\label{sec:sim_details}
The simulations of the \ac{QAOA} circuit for the \ac{TFIM} are done in the free fermion picture yielding a quadratic scaling in $N$ for the cost function evaluation. 
The circuits including $L_y$ layers and for the \ac{XXZM} do not obey the same symmetries and therefore are implemented as a full circuit simulation using \textsc{ProjectQ} \cite{ProjectQ}. 
The depth of the \ac{QAOA} circuit for the \ac{TFIM} is fixed to the smallest value containing the exact ground state $p=N/2$, which gives us $N$ variational parameters with one added per $L_y$ in the second main experiment.
For the \ac{XXZ} model we choose $p=N$ resulting in $3N$ variational parameters. 
All circuit simulations are performed exactly, \ie without noise or sampling.
Furthermore we use the \textsc{SciPy} implementation of the \ac{BFGS} algorithm and in-house routines for \ac{ADAM} and \ac{NatGrad} \cite{SciPy}.
All variational parameters are initialized uniformly i.i.d. over the interval $[0.0001,0.05]$ as this corresponds to initializing the circuit close to the identity and symmetric randomization around $0$ has shown slightly worse performance in our experiments.

We bound the \ac{BFGS} optimization to one period of the rotation parameters as this improves the line search efficiency and found only a small dependence on the position of the interval.
For the \ac{ADAM} optimizer we fixed $\gb_1=0.9$, $\gb_2=0.999$ and $\ge=10^{-7}$ and vary $\eta$ in $[0.005,0.5]$ trying to build heuristics for the particular problems. 
We found non-trivial behaviour of \ac{ADAM} \wrt the learning rate, observing a strong influence on the optimization duration, for details see sec.~\ref{sec:res_tfi_qaoa}.
Furthermore, an increased regularization constant $\ge$ did not yield any improvements of \ac{ADAM}.
For \ac{NatGrad} we fix the Tikhonov regularization to $\ge_T=10^{-4}$ and use learning rates of $0.5$, $0.05$ and $0.1$. 
Employing block diagonal approximations to the Fubini-Study matrix as suggested in \cite{Stokes_Carleo_19} was not successful due to long-range correlations between the variational parameters in the circuit.

\section{Conclusion}\label{sec:conclusion} 

Our first main result shows that the \ac{BFGS} optimizer, while quick and reliably for small systems, has an increased chance getting stuck in local minima already in medium sized \acp{VQE}, in the range of present day and near future \ac{NISQ} devices.
This may be surprising as it has access to non-local information due to its line search subroutine. 
We suspect that this aspect of the algorithm becomes less helpful for finding a global minimum because of its sparsity in high-dimensional parameter spaces.

The \ac{ADAM} optimizer on the other hand is able to find global minima also in larger parameter spaces (up to $42$) for suitably small learning rates but this comes at the cost of a quickly increasing number of epochs to complete the optimization.
In particular we observed two effects of the learning rate $\eta$ on the runtime of \ac{ADAM}:
On the one hand, there is a threshold size of the parameter space that depends on $\eta$ above which the epoch count rapidly increases, which means that a small enough value of the learning rate is essential to avoid extremely long runtimes.
On the other hand, the optimization duration for sizes below the threshold is significantly increased when reducing $\eta$ such that it is undesirable to choose it smaller than strictly necessary.
It thus appears that tedious hyperparameter tuning may be necessary to balanced these two effects.

The \ac{NatGrad} optimizer recently proposed for \acp{VQE} shows very reliable convergence to a global minimum for all tested system sizes within fewer epochs but at high per-epoch cost.
We found that Tikhonov regularization can fix the problem of getting lost in barren plateaus even after a suitable initialization.
This makes the algorithm a promising, although expensive, candidate for the optimization of future \acp{VQE}.
The increased cost for determining the Fubini matrix at each step have a particularly strong effect on the estimated quantum runtime for spin chain systems, such that for other systems with more favourable scaling \ac{NatGrad} might not only be more reliable but additionally exhibit competitive cost.

Our second main experiment treats overparametrization in \ac{VQE} ansatz classes using the example of additional rotation gates that break the symmetry of the Hamiltonian.
The \ac{BFGS} optimizer fails to find a global minimum in some instances even for small systems and in general exhibits a strongly fluctuating performance which decreases with the number of additional gate layers.
The simulation cost restricted the maximal system size for this second experiment but there is no reason to assume that a stronger overparametrization with more symmetry breaking layers would resolve these problems.

Also \ac{ADAM} showed strong susceptibility to the additional degrees of freedom. 
Beyond the implications on applications, this is interesting because overparametrization is heavily used in machine learning to make the cost function tractable for optimizers like \ac{ADAM} and we therefore appear to observe a fundamental difference between classical machine learning and \acp{VQE}.

Finally, \ac{NatGrad} showed some failed optimization runs for selected system sizes as well but mostly remained successful even for multiple additional gate layers. 
It therefore rewards its increased cost per epoch with higher success rates and is the only tested optimization strategy that showed resilience to both, big search spaces and local minima caused by overparametrization.

The extension of our analysis to the \ac{XXZM} confirmed the problems of the \ac{BFGS} optimizer with big search spaces and the rapid runtime growth for \ac{ADAM}. 
\ac{NatGrad} performed less reliably on the \ac{XXZM} and the per-epoch cost dominate the reduced number of epochs. 
The convergence issues might be either due to local minima or optimization interrupts based on small improvements with a series of updates, where preliminary insights suggest that the latter is the case and that \ac{NatGrad} could be improved by tailoring it to \acp{VQE}. 

Our investigations have shown that \ac{NatGrad} might enable \acp{VQE} to solve more complex and bigger problems as it performs well on a test model with challenges representative of those in potential future applications of \acp{VQE}.
Caution is in order, however, when generalizing this result to other models as we saw in the case of the \ac{XXZM}.

\section{Acknowledgements}
We would like to thank Chae-Yeun Park, David Gross, Gian-Luca Anselmetti and Thorben Frank for helpful discussions. 
We acknowledge funding by the Deutsche Forschungsgemeinschaft (DFG, German Research Foundation) under Germany's Excellence Strategy – Cluster of Excellence Matter and Light for Quantum Computing (ML4Q) EXC 2004/1 – 390534769.
The authors would like to thank Covestro Deutschland AG, Kaiser Wilhelm Allee 60, 51373 Leverkusen, for the support with computational resources.
The work was conducted while all three authors were affiliated with the Institute for Theoretical Physics of the University of Cologne.

\bibliography{lib.bib}

\begin{appendix}
\section{Exact solution of the TFIM}\label{sec:app_tfi}
Here we derive the analytic solution of the \ac{TFIM} by mapping it to non-interacting fermions, also see \cite{Wang_Rieffel_19}.
We start with the linear combinations $a_k\coloneqq \frac{1}{2}(Z^{(k)}+iY^{(k)})$ which fulfil
\begin{equation}\label{eq.comp:pauli_of_a}
X^{(k)} = 2a_k\dag a_k-1\;,\; Z^{(k)} = a_k\dag + a_k
\end{equation} 
and map them to the operators
\begin{equation}
b_k \coloneqq \p{l}{1}{k-1}\cN_l\;a_k 
\; ,\; \cN_l\coloneqq \exp\left[i\pi a_l\dag a_l\right] 
\end{equation}
which satisfy fermionic anticommutation relations:
\begin{equation}\label{eq.def:b_of_a}
\{b_k\dag,b_l\}=\gd_{kl}\;,\;\{b_k,b_l\}=\{b_k\dag,b_l\dag\}=0.
\end{equation}
For the transformation of the Hamiltonians $H_S$ and $H_B$, which comprise both the \ac{TFIM} Hamiltonian and the generators for the unitaries in the \ac{QAOA} ansatz, note that
\begin{align}
\cN_l^2&=\unit\;,\;\cN_l\dag=\cN_l=\cN_l^{-1}\\
\cN_kb_k&=b_k\;,\;\cN_kb_k\dag=-b_k\dag . \label{eq.comp:N_at_b}
\end{align}
Using eqn.~(\ref{eq.comp:pauli_of_a}) and the above properties the transformed Hamiltonians read
\begin{align}
H_S &= -\left[\s{k}{1}{N-1} (b_k\dag-b_k)b_{k+1}\dag \right.\\&\left.\hspace{1.75cm}- (b_N\dag-b_N)b_{1}\dag\;\mathcal{G}\right]+h.c.\;,\NN\\
H_B &= -t\s{k}{1}{N}2b_k\dag b_k-1
\end{align}
where we denote by $\cG\coloneqq\p{l}{1}{N}\cN_l$ the gauge factor in the term generated by the periodic boundary conditions and the non-local transformation eqn.~(\ref{eq.def:b_of_a}) which also has a reversed sign.
$\cG$ interacts with the initial state of the \ac{QAOA} ansatz $\ket{\bar{\psi}}$ and the Hamiltonian terms in the following way: 
\begin{align}
\cG\ket{\bar{\psi}}&=\exp\left[ \frac{i\pi}{2} \left(-\frac{1}{t}H_B+N\right)\right]\ket{+}^{\otimes N}=e^{i\pi N}\ket{\bar{\psi}},\\
[\cG,H_B] &= 0=[\cG,H_S]
\end{align}
where we used the ground state energy $-tN$ of $H_B$ and eqn.~(\ref{eq.comp:N_at_b}).
This means that the reversed sign is cancelled for odd $N$.
Therefore we introduce an additional phase via the transformation
\begin{align}
c_k&\coloneqq e^{ik\nu}b_k\; ,\; \nu\coloneqq\cas{\pi/N}{N\text{ even}}{0}{N\text{ odd}},\\
H_S &= -\left[\s{k}{1}{N} \;e^{i\nu}\left( c_k\dag e^{i2k\nu} -c_k  \right)c_{k+1}\dag \right] +h.c.\;, \\
H_B &= -t\s{k}{1}{N} 2c_k\dag c_k-1
\end{align}
where we defined $\nu$ such that the result holds for both odd and even $N$.
The last mapping we perform is a Fourier transformation with shifted momenta:
\begin{align}
d_q &\coloneqq \frac{1}{\sqrt{N}}\s{k}{1}{N}e^{2\pi i(q-1)k/N}c_k\;,\\ 
H_S &= -\left[\s{q}{1}{N} \; e^{-i\ga_q}d_q\dag d_{-q}\dag 
-e^{i\ga_q} d_q d_{q}\dag\right] +h.c.\;,\\
H_B &= t\s{q}{1}{N}2d_q\dag d_q-1 
\end{align}
with mode-dependent angles and relabeled Fourier modes
\begin{align}
\ga_q&\coloneqq \cas{(2q-1)\pi/N}{N\text{ even}}{2q\pi/N}{N\text{ odd}}\\
d_{-q}&\coloneqq \cas{d_{N+1-q}}{N\text{ even}}{d_{N+2-q}}{N\text{ odd}}.
\end{align}
We finally can split up the sums, recollect the terms corresponding to the pairs $\{d_q,d_{-q}\}$ and rewrite the Hamiltonians in a fermionic operator basis:
\begin{align}
H_S &= H'_{S}-2\left[\s{q}{1}{r} \; \cos\alpha_q \left(d_{q}\dag d_q-d_{-q} d_{-q}\dag\right) \right.\NN\\
&\left.\hspace{2.25cm}-i\sin\alpha_q\left(d_q\dag d_{-q}\dag-d_{-q}d_{q}\right) \right]\\
&=-2\s{q}{1}{r}
\left(\begin{matrix}d_q\dag & d_{-q}\end{matrix}\right)
\left(\begin{matrix}\cos\ga_q & -i\sin\ga_q \\ i\sin\ga_q & -\cos\ga_q \end{matrix}\right)
\left(\begin{matrix}d_q \\ d_{-q}\dag\end{matrix}\right)\NN\\
&\quad+H'_{S},\\
H_B&=H'_{B}-2t\s{q}{1}{r}d_q\dag d_q -d_{-q}d_{-q}\dag\\
&=H'_{B}-2t\s{q}{1}{r}
\left(\begin{matrix}d_q\dag & d_{-q}\end{matrix}\right)
\left(\begin{matrix}1 & 0 \\ 0 & -1 \end{matrix}\right)
\left(\begin{matrix}d_q \\ d_{-q}\dag\end{matrix}\right),
\end{align}
where $H'_{B}=H'_{S}=0$ and $H_{B}/t=H'_{S}=-1$ for even and odd $N$ respectively, using $d_1\dag d_1\ket{\bar{\psi}}=1$ and $d_1d_1\dag\ket{\bar{\psi}}=0$ for the odd case.

In this shape the simple structure of the model becomes apparent as we identify $r$ pairs of fermionic modes in momentum space which interact within but not between the pairs.
The Hamiltonian can thus be written as a direct sum
\begin{equation}
	\begin{split}
H_\text{TFI}=&-2\bigoplus_{q=1}^r (t+\cos\ga_q) Z +\sin\ga_q Y \\
&-(1+t)(N-2r).
\end{split}
\end{equation}
Due to the fact that $H_B$ and $H_S$ not only constitute $H_\text{TFI}$ but also generate the (modified) \ac{QAOA} ansatz, the simulation of the circuit can be carried out on a $2r$-dimensional space that decomposes into the direct sum above. 
On the Bloch spheres of the free fermions the two time evolution operators $e^{-i\gt H_S}$ and $e^{-i\gf H_B}$ correspond to rotations about the individual axes $e_k=(0,\sin\ga_q,\cos\ga_q)$ and the z-axis respectively.
Furthermore we can manually solve for the ground state of the \ac{TFIM} by computing the ground state in each subspace individually:
\begin{align}
E_0&=E'-2\s{q}{1}{r} E_q,\;\ket{\psi_0}=\bigoplus_{q=1}^{r} \ket{\psi_{q,0}},\\
E_{q}&=\sqrt{1+t^2+2t\cos\ga_q},\\
\ket{\psi_{q,0}}&=\frac{1}{\sqrt{2E_q(E_q-\cos\ga_q-t)}} \bmat{c} i\sin\ga_q \\ E_q-\cos\ga_q-t \emat
\end{align}
where $E'$ is the eigenvalue of $H'_B+H'_S$.

\section{Learning rate influence on \ac{ADAM} performance}\label{sec:app_eta_Adam}

\begin{figure}
\flushleft
\vspace{0.25cm}
\includegraphics[width=0.48\textwidth]{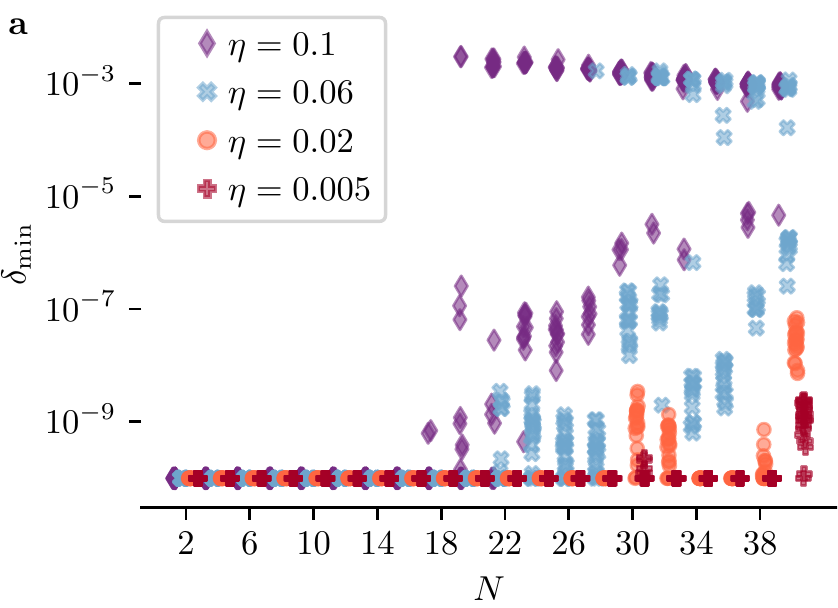}

\vspace{0.25cm}
\includegraphics[width=0.48\textwidth]{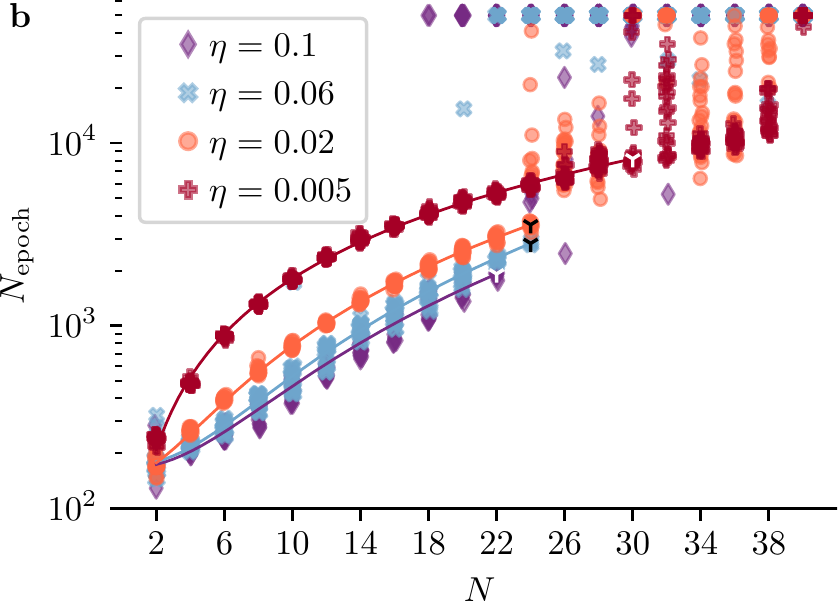}
\caption{Minimal attained relative errors $\gd_\text{min}$ and epoch count $N_\text{epoch}$ for the \ac{ADAM} optimizer initialized at $20$ distinct points close to zero and with different learning rates $\eta$. 
(a) The threshold size beyond which \ac{ADAM} fails can be shifted by reducing $\eta$, delaying local convergence to bigger systems.
(b) The shown fits are based on \textit{filtered} data in order to determine the apparent scaling for small system sizes and thus do \textit{not} aim at describing the entire data.
The biggest system size partially included in the fit is marked. 
For the shown learning rates in descending order, we obtain the exponents $2.3$, $2.3$, $1.9$ and $1.4$ but prefactors $1.8$, $1.9$, $7.3$ and $74.7$. 
\label{fig:tfi_eta_adam}}
\end{figure}

In order to evaluate the systematically large optimization durations of the \ac{ADAM} optimizer for the \ac{QAOA} circuit of the \ac{TFIM}, we tested it at multiple learning rates from the interval $[0.005,0.1]$ observing a major influence on the runtime, see fig.~\ref{fig:tfi_eta_adam}.
For a given learning rate $\eta$, the required number of epochs grows polynomially with the system size up to a size $N^*$ above which \ac{ADAM} takes much longer, exceeding the budget of $5\cdot 10^4$ iterations.
In this second phase we find the optimizer to require excessively many iterations both when succeeding and when getting stuck in a local minimum (see \eg $\eta=0.06$), which prevents us from systematically distinguishing the two cases before convergence.
The observed transition point $N^*(\eta)$ can be shifted towards bigger system sizes by decreasing the learning rate, \ie $N^*(\eta)$ is monotonically decreasing.
Meanwhile, reducing $\eta$ increases the epoch count significantly for smaller system sizes without disrupting the convergence as is expected for well-behaved systems.
Even though the scaling exponent is smaller for lower learning rates the optimization requires more iterations which is due to a large prefactor, such that the cost are increased for all system sizes before the jump.
The observed dependencies of the runtime on $\eta$ result in a system size dependent optimal learning rate which trades off the systematically increased epoch counts for small $\eta$ against the position of the jump in optimization duration.
This demonstrates that heuristics for \ac{ADAM} are needed in order to achieve systematic global optimization and that the required numer of optimization steps can be unpredictably large depending on the hyperparameters.

\end{appendix}

\end{document}